\documentclass{webofc}
\usepackage[varg]{txfonts}   
\usepackage{orcidlink}
\hypersetup{hidelinks}

\begin{document}
%
\title{Search for anomalous chiral effects in heavy-ion collisions with ALICE}
%
%

\author{\firstname{Chun-Zheng}
\lastname{Wang\orcidlink{0000-0001-5383-0970} (for the ALICE Collaboration)}\inst{1}\fnsep\thanks{\email{chunzheng.wang@cern.ch}}
}

\institute{Key Laboratory of Nuclear Physics and Ion-beam Application (MOE), Institute of Modern Physics, Fudan University, 200433, Shanghai, China
}
\abstract{%
The interplay between the chiral anomaly and the strong magnetic or vortical fields created in noncentral heavy-ion collisions can lead to various anomalous chiral effects in the quark--gluon plasma, including the chiral magnetic effect (CME), the chiral magnetic wave (CMW), and the chiral vortical effect (CVE). In this proceeding, recent ALICE measurements of these effects are summarized. Utilizing Event Shape Engineering, fractions of CME and CMW signals extracted in Pb--Pb collisions at $\sqrt{s_{\rm{NN}}} = 5.02$ TeV are consistent with zero within uncertainties. The CVE is studied using azimuthal correlations between baryon pair $\Lambda$--p with $\Lambda$--h and h--h as reference. These measurements provide new insights into the experimental search for anomalous chiral effects in heavy-ion collisions.
}
\maketitle
\section{Introduction}
\label{intro}
 Within the quark--gluon plasma (QGP) created by the relativistic heavy-ion collisions, QCD vacuum fluctuations may generate localized regions with chiral and baryonic imbalances, essentially resulting in finite chiral and baryonic chemical potentials. Under such conditions, the strong magnetic and vortical fields produced by these collisions could prompt various anomalous chiral effects \cite{ThyCME2,ThyCMW,ThyCVE2} such as the chiral magnetic effect (CME) and chiral magnetic wave (CMW) driven by magnetic fields, as well as the chiral vortical effect (CVE) induced by vortical fields. The study of these phenomena is of fundamental significance since their existence may reveal the topological structure of vacuum gauge fields, as well as the possible local violation of P and/or CP symmetries in strong interactions.
 
 In measurements of CME, which are expected to introduce an electric dipole moment in the QGP fireball, the three-particle correlator $\gamma = \langle \cos(\varphi_a+\varphi_b - 2\Psi_{2}) \rangle$ is widely applied \cite{ObvCME}. Here, $\varphi_a$ and $\varphi_b$ correspond to azimuthal angles of two different charged particles, while $\Psi_{2}$, representing the second order symmetry plane, is nearly perpendicular to the magnetic field. This orientation of $\Psi_{2}$ makes $\gamma$ sensitive to potential CME signals. The two-particle correlator $\delta = \langle \cos(\varphi_a-\varphi_b) \rangle$ is sometimes calculated simultaneously to study the background. One can take the difference between correlators of opposite-sign charge pairs ($\gamma_{\rm{OS}}$ and $\delta_{\rm{OS}}$) and same-sign charge pairs ($\gamma_{\rm{SS}}$ and $\delta_{\rm{SS}}$), denoted as $\Delta\gamma$ and $\Delta\delta$, to eliminate charge-independent correlation background, such as transverse momentum conservation. The CMW introduces an electric quadrupole moment in the QGP fireball and its analysis involves extracting the linear slope $r$ between the event-by-event charge asymmetry $A_\mathrm{ch} = (N^{+} - N^{-})/(N^{+} + N^{-}$) (with $N^{+}$ and $N^{-}$ denoting the number of positive and negative particles, respectively) and the difference between the elliptic flow of negative and positive particles $v_{2}^{-} - v_{2}^{+}$. Alternatively, measuring CMW can also be achieved by calculating the covariance between $A_\mathrm{ch}$ and $v_{2}^{-} - v_{2}^{+}$, which has the advantage of not requiring efficiency corrections \cite{ObvCMW}. In the case of CVE measurements, analysis process is very similar to CME. However, given that the polarization of quarks by the vortical field is related to their baryon number, one should calculate the correlators between baryon pairs instead of charge pairs.
 
\begin{figure}
\centering
\includegraphics[width=6.3cm,clip]{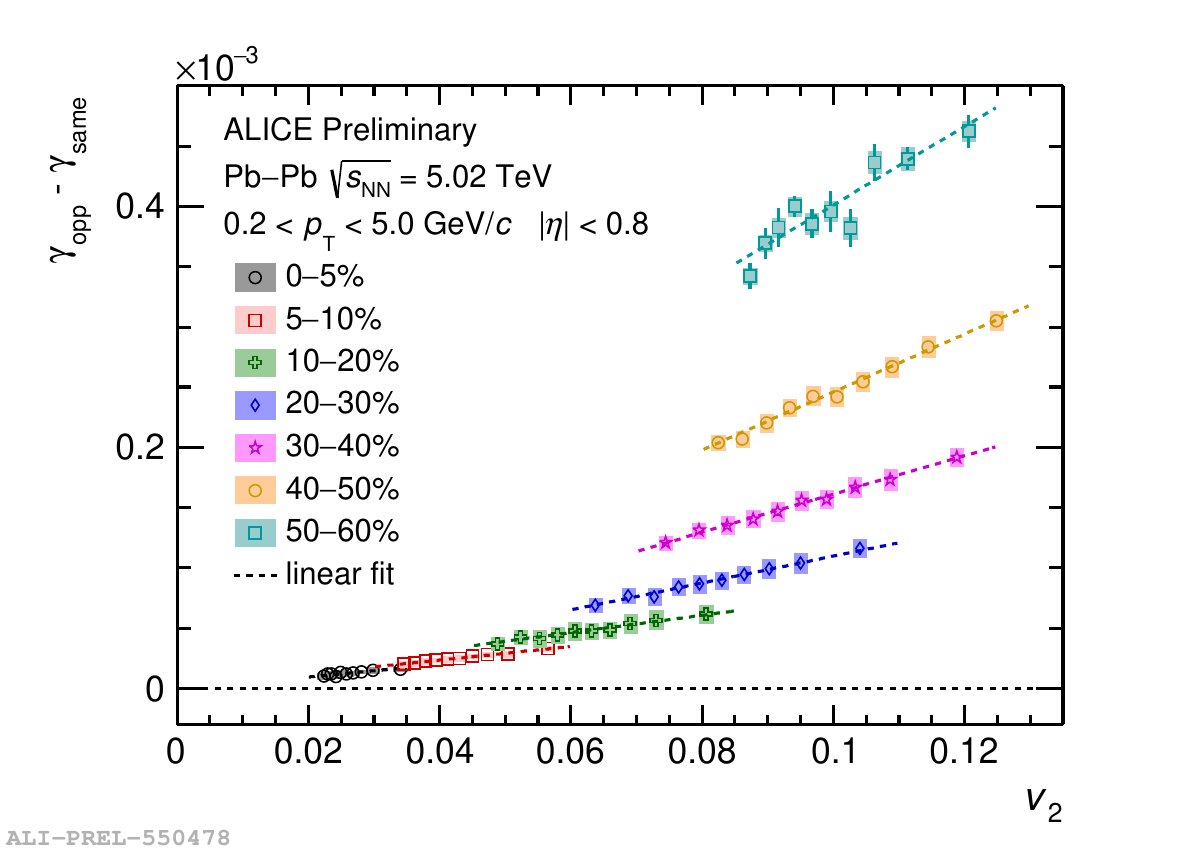}
\includegraphics[width=6.3cm,clip]{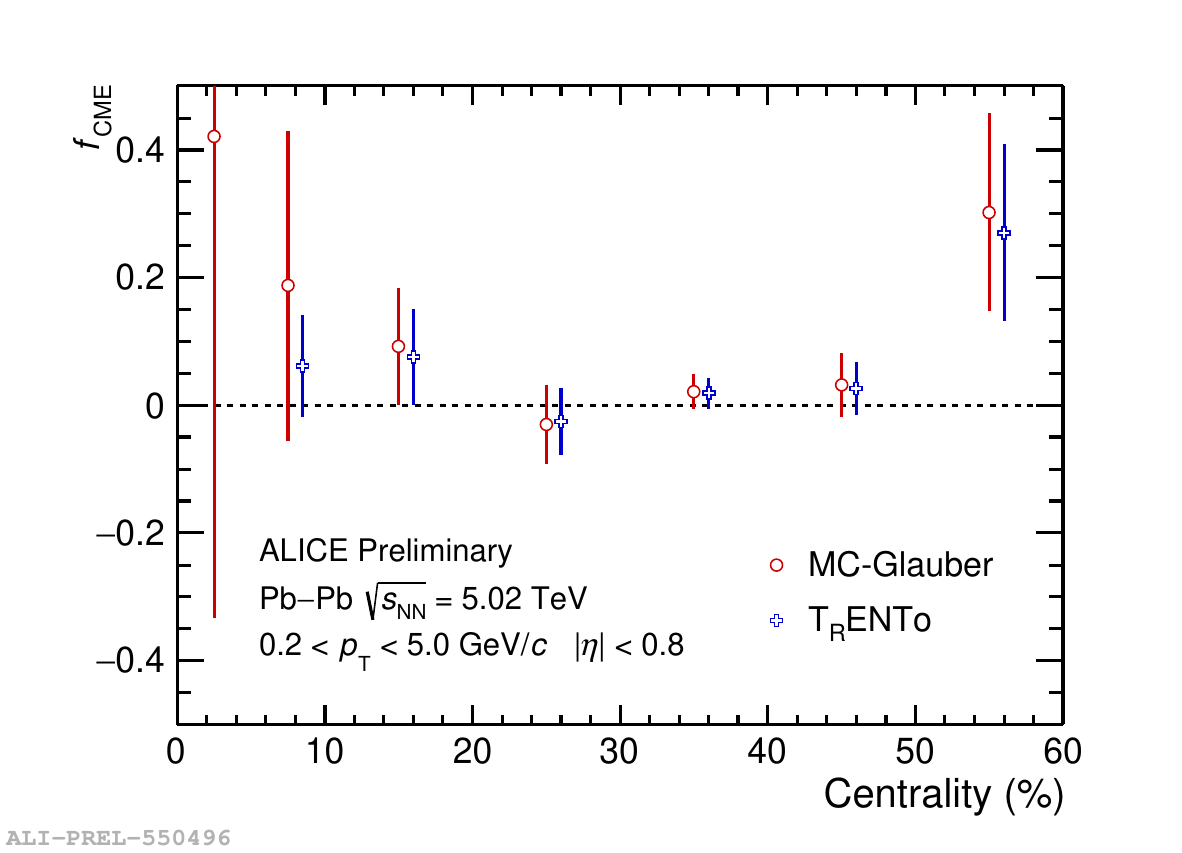}
\vspace*{-0.3cm}       
\caption{Left: Difference between $\gamma$  correlator of opposite and same charge pair as a function of $v_{2}$ in various centrality classes. Right: Centrality dependence of the CME fraction extracted from the slope parameter of fits to data and MC-Glauber and $\text{T}_\text{R}\text{ENTo}$ models.}
\label{fig-1}       
\end{figure}
 
 The ALICE Collaboration has conducted measurements of CME and CMW using the observables previously mentioned. In these measurements, clear separations between $\gamma_{\rm{OS}}$ and $\gamma_{\rm{SS}}$ for CME, and robust relationships between $v_{2}^{-} - v_{2}^{+}$ and $A_\mathrm{ch}$ for CMW have been observed \cite{ExpALICECME,ExpALICECMW}, seemingly bearing out the theoretical predictions. However, extensive research indicates that a significant part of the observed values may be due to background effects, especially the local charge conservation (LCC) entwined with collective flow. For example, the recent work \cite{BkgCMECMW} has shown that the measurements of CME and CMW with ALICE can be uniformly explained by a set of parameters within a blast-wave model incorporating LCC. Therefore, extracting the fraction of real signals arising from chiral anomalies and understand the background represent crucial aspects of current research. In this proceeding, the recent measurements on CME, CMW, and CVE from ALICE are summarized.

\section{Recent results from ALICE}
\label{sec-Results}

\subsection{Extraction of CME fraction}
\label{subsec-CME}

Event shape engineering (ESE) \cite{ESE} is employed to extract the fraction of CME $f_{\rm{CME}}$ in Pb--Pb collisions at $\sqrt{s_{\rm{NN}}} = 5.02$ TeV in this analysis. The left panel of Fig. \ref{fig-1} shows the $v_{2}$ dependence of the $\Delta\gamma$ (i.e. $\gamma_{\rm{opp}} - \gamma_{\rm{same}}$)
across different centrality intervals. Two Monte Carlo models, MC-Glauber \cite{MC1} and $\text{T}_\text{R}\text{ENTo}$ \cite{MC2}, are utilized in this study to evaluate the expected dependence of the CME signal on $v_{2}$. The CME signal is assumed to be proportional to $\langle |B|^2 \cos(2(\Psi_{\rm{B}} - \Psi_{2}))\rangle$, where $B$ and $\Psi_{B}$ represent the magnitude and direction of the magnetic field, respectively. To distinguish the CME signal from background, the measured $\Delta\gamma$ and the estimated $\langle |B|^2 \cos(2(\Psi_{\rm{B}} - \Psi_{2}))\rangle$ are both fitted with a linear function $f(v_2) = p_0 \left [ 1+p_{1}(v_2-\langle v_{2} \rangle)/{\langle v_{2} \rangle} \right ]$, where $p_0$ represents the overall scale and $p_1$ is related to the potential CME signal. It is assumed that the background scales linearly with $v_2$ and in a pure background scenario $p_1$ is equal to unity. The $f_{\rm{CME}}$ extracted by the formula $p_{1,\rm{data}} = f_{\rm{CME}}p_{1,\rm{MC}} + (1-f_{\rm{CME}})$ in the different centrality intervals is shown in the right panel of Fig. \ref{fig-1}.

After combining the results from the 5--60\% centrality interval, the $f_{\rm{CME}}$ extracted is 0.028 ± 0.021 (with MC-Glauber) or 0.025 ± 0.018 (with $\text{T}_\text{R}\text{ENTo}$). These correspond to upper limits of 6.4\% and 5.5\%, respectively, at a 95\% confidence level (C.L.). Taking advantage of a larger data sample in ALICE Run 2 compared to Run 1, this recent analysis shows significantly reduced uncertainties compared to Ref. \cite{ExpESECMERun1}, which used the same method to extract $f_{\rm{CME}}$ in Pb--Pb collisions at $\sqrt{s_{\rm{NN}}} = 2.76$ TeV. Additionally, an analysis employing a two-component approach in Xe--Xe collisions at $\sqrt{s_{\rm{NN}}} = 5.44$ TeV is presented in Ref. \cite{ExpESEXeXe}.

\begin{figure}
\centering
\includegraphics[width=6.cm,clip]{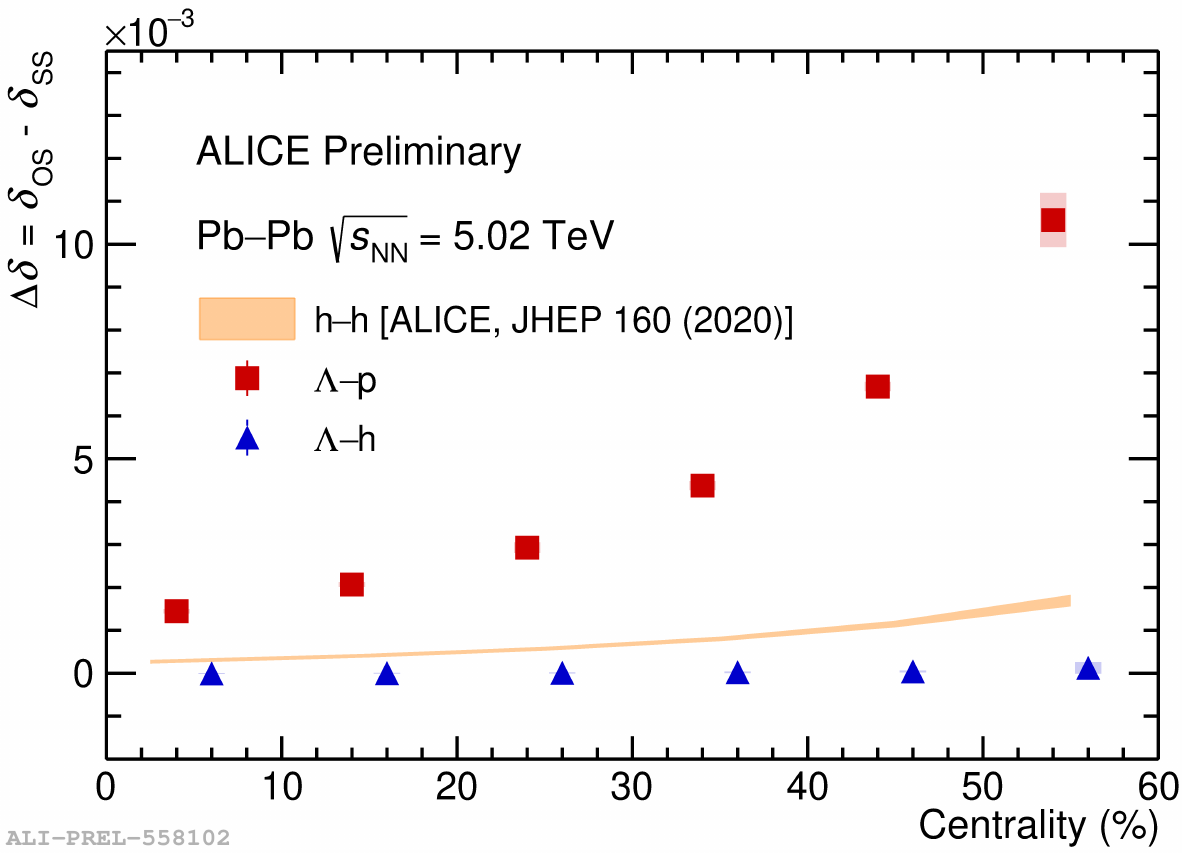}
\includegraphics[width=6.cm,clip]{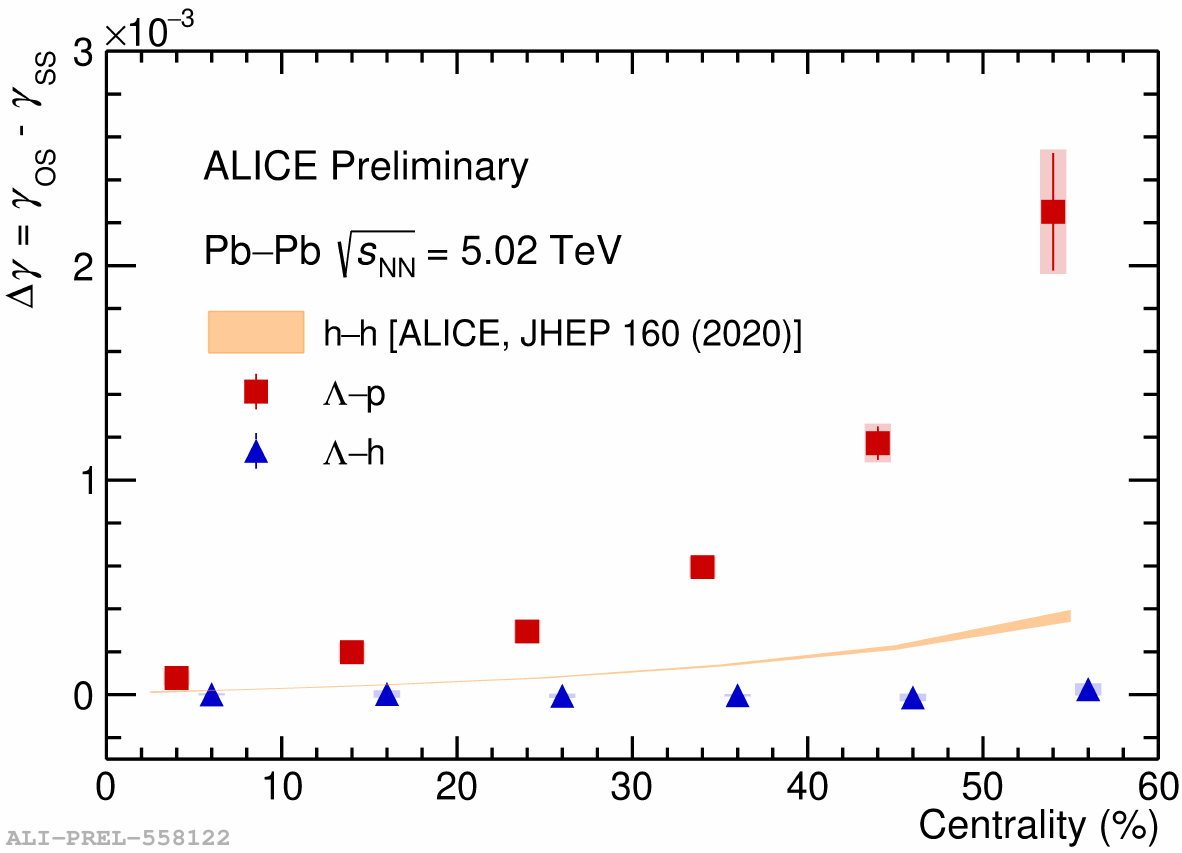}
\vspace*{-0.3cm}       
\caption{Centrality dependence of $\Delta\delta$ (left panel) and $\Delta\gamma$ (right panel) between the opposite and same sign pairs $\Lambda$--p, $\Lambda$--h, and h--h.}
\label{fig-2}       
\end{figure}
\label{subsec-CVE}

\subsection{Extraction of the CMW fraction}
\label{subsec-CMW}
The ESE method is also used to separate LCC background contributions from the potential CMW signal, as proposed in Ref. \cite{ESECMW}. For the 10--60\% centrality interval, the CMW fraction is 0.081 $\pm$ 0.055 with the upper limits of 26\% at a 95\% C.L. and 38\% at a 99.7\% C.L. Detailed information on this analysis can be found in Ref. \cite{ExpESECMW}.

\subsection{Search for the CVE}
\label{subsec:CVE}

\begin{figure}
\centering
\includegraphics[width=6.cm,clip]{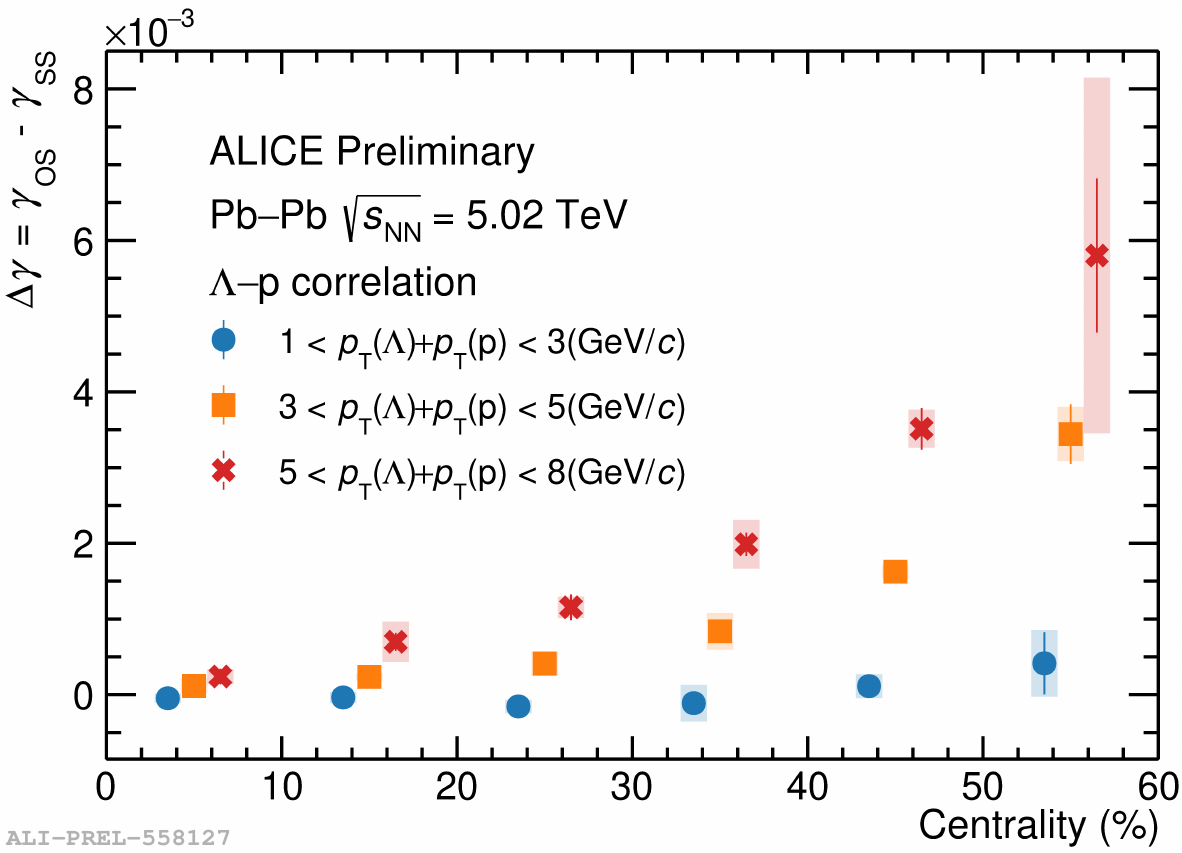}
\includegraphics[width=6.cm,clip]{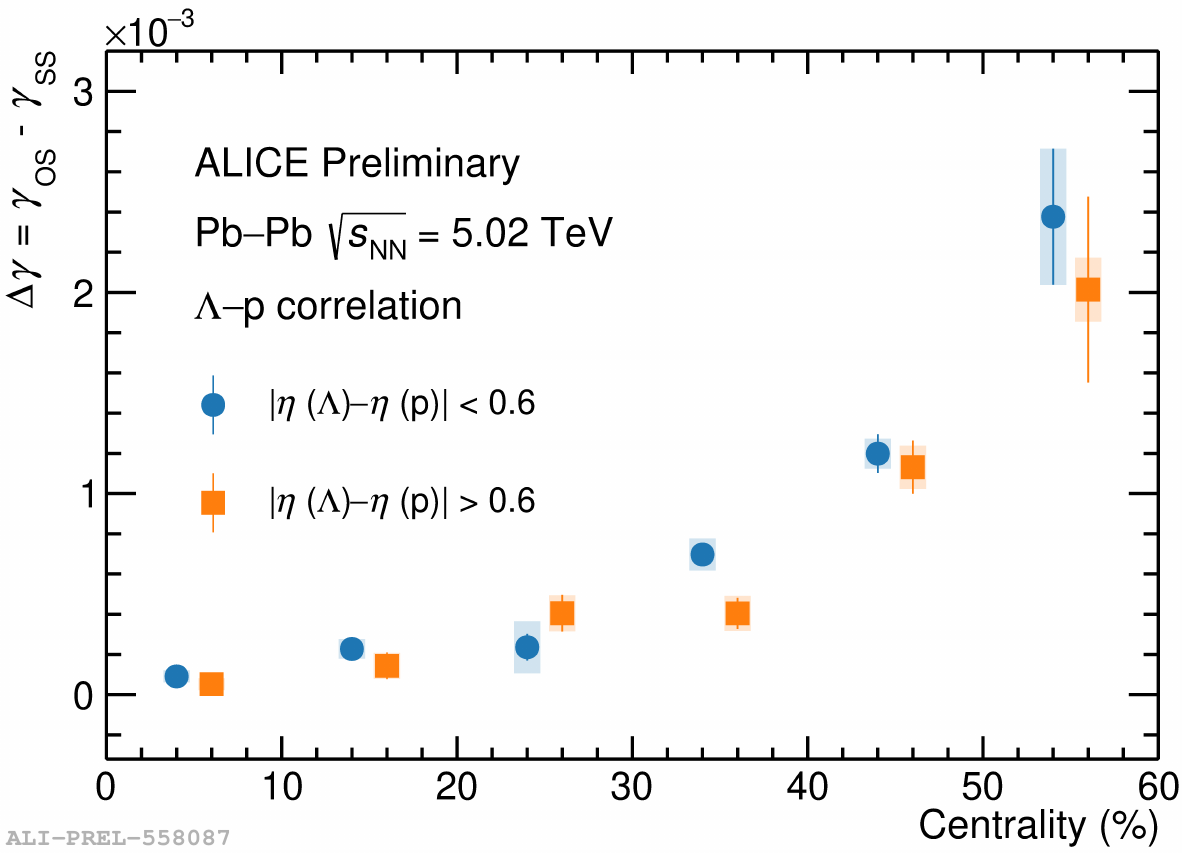}
\vspace*{-0.3cm}       
\caption{Centrality dependence of $\Delta\gamma$ between the opposite and same sign pairs $\Lambda$--p in the different $\sum{p_{\rm{T}} = p_{\rm{T}}(\Lambda) + p_{\rm{T}}(\text{p})}$ intervals (left panel) and $\Delta\eta = |\eta(\Lambda) - \eta(\text{p})|$ intervals (right panel).}
\label{fig-3}       
\end{figure}

In the analysis of CVE, $\delta$ and $\gamma$ of baryon pair $\Lambda$--p are investigated with $\Lambda$--h, and h--h as reference. The neutral baryon $\Lambda$ is chosen to avoid the influence of charge-dependent effects (like CME). Figure \ref{fig-2} presents the centrality dependence of the $\Delta\delta$ and $\Delta\gamma$ in the left and right panels, respectively. A notable $\delta$ and $\gamma$ separation is observed in $\Lambda$--p, which is approximately 10 times larger than that in h--h. However, both $\delta$ and $\gamma$ separation in $\Lambda$--h are almost negligible. These results suggest a strong azimuthal correlation specifically in $\Lambda$--p pairs, which partially align with predictions for CVE, but further analysis is required to assess potential background.

The analysis was also performed by varying the kinematic ranges. The centrality dependence of $\Delta\gamma$ in the different transverse momentum intervals $\sum{p_{\rm{T}} = p_{\rm{T}}(\Lambda) + p_{\rm{T}}(\text{p})}$ is shown in the left panel of Fig \ref{fig-3}. The results indicate that larger $\sum{p_{\rm{T}}}$ yields larger $\Delta\gamma$. The right panel of Fig. \ref{fig-3} displays the centrality dependence of $\Delta\delta$ and $\Delta\gamma$ under two sets of $\Delta\eta = |\eta(\Lambda) - \eta(\text{p})|$, showing that the impact of $\Delta\eta$ on $\Delta\gamma$ is not very significant within uncertainties. These results may suggest that the underlying background has a strong correlation with the transverse momentum of the particles, but is not much related to the pseudorapidity difference between the particles. This will help in establishing models to analyze the background.

\subsection{Summary}

Anomalous chiral effects, including CME, CMW, and CVE, have been studied in Pb--Pb collisions at $\sqrt{s_{\rm{NN}}} = 5.02$ TeV using the ALICE detector. The ESE method has been successfully applied to extract limits to potential CME and CMW signals. The fractions of both CME and CMW are consistent with zero within uncertainties. Furthermore, the first measurement of CVE performed by ALICE have revealed non-trivial behaviors in the $\delta$ and $\gamma$ correlators of the baryon pair $\Lambda$--p, which also exhibit a hierarchy across different kinematic regions. The background of CVE measurement still requires comprehensive theoretical study to fully understand these results and disentangle the CVE signal from the background.

This work was supported by the National Key Research and Development Program of China (No. 2018YFE0104600) and the National Natural Science Foundation of China (Nos. 11975078, 12061141008, 12322508).

%
%
%

\end{document}